\documentclass[a4paper,conference]{IEEEtran}
\IEEEoverridecommandlockouts
\usepackage{cite}
\usepackage{amsmath,amssymb,amsfonts}
\usepackage{algorithmic}
\usepackage{graphicx}
\usepackage{textcomp}
\usepackage[nolist, nohyperlinks]{acronym}
\usepackage{xcolor}
\usepackage{orcidlink}
\def\BibTeX{{\rm B\kern-.05em{\sc i\kern-.025em b}\kern-.08em
    T\kern-.1667em\lower.7ex\hbox{E}\kern-.125emX}}
\usepackage[ruled,norelsize]{algorithm2e}
    
\begin{document}

\makeatletter
\newcommand{\linebreakand}{
    \end{@IEEEauthorhalign}
    \hfill\mbox{}\par
    \mbox{}\hfill\begin{@IEEEauthorhalign}
}
\newcommand{\removelatexerror}{\let\@latex@error\@gobble}
\makeatother

\begin{acronym}
    \acro{GEO}{Geostationary Orbit}
    \acro{LEO}{Low Earth Orbit}
    \acro{ISL}[ISL]{Intersatellite Link}
    \acro{ESL}{Earth-Satellite Link}
    \acro{KPI}{Key Performance Indicator}
    \acro{SCN}{Satellite Constellation Network}
    \acro{SDN}[SDN]{Software Defined Networking}
    \acro{CPP}[CPP]{Controller Placement Problem}
    \acro{SR}[SR]{Segment Routing}
    \acro{QoS}{Quality of Service}
    \acro{RTT}{Round-Trip Time}
    \acro{SLA}{Service Level Agreement}
    \acro{SLR}{Service Level Requirement}
    \acro{SPF}{Shortest Path First}
    \acro{ML}{Machine Learning}
    \acro{DRL}{Deep Reinforcement Learning}
    \acro{DQN}{Deep Q-Network}
    \acro{UT}{User Terminal}
    \acro{GW}{Gateway station}
    \acro{GEVR}{God's Eye View Routing}
    \acro{SSPF}{Source-routed Shortest Path First}
    \acro{ELB}{Explicit Load Balancing}
    \acro{CPP}{Controller Placement Problem}
    \acro{TLR}{Traffic-Light-based intelligent Routing strategy}
    \acro{DRA}{Datagram Routing Algorithm}
    \acro{IDLB}{Independent Distributed Load-balanced routing}
\end{acronym}

\title{Distributed SDN-based Load-balanced Routing for Low Earth Orbit Satellite Constellation Networks\\

\thanks{This work has been supported by the Advanced Research in Telecommunications Systems Programme of the European Space Agency (ESA), activity code 3A.117. Responsibility for the contents of this publication rests with the authors.}
}

\author{
\IEEEauthorblockN{Manuel M. H. Roth, Hartmut Brandt, Hermann Bischl}
\IEEEauthorblockA{\textit{Institute of Communications and Navigation} \\
German Aerospace Center (DLR), Weßling, Germany\\
\{manuel.roth \orcidlink{0000-0001-7878-1204}, hartmut.brandt, hermann.bischl\}@dlr.de}
}

\maketitle

\begin{abstract}
With the current trend towards low Earth orbit mega-constellations with inter-satellite links, efficient routing in such highly dynamic space-borne networks is becoming increasingly important. Due to the distinct network topology, specifically tailored solutions are required. Firstly, the relative movement of the constellation causes frequent handover events between the satellites and the terminals on ground. Furthermore, unevenly distributed traffic demands lead to geographical hot spots. The physical size of the network also implies significant propagation delays. Therefore, monitoring the dynamic topology changes and link loads on a network-wide basis for routing purposes is typically impractical with massive signaling overhead. To address these issues, we propose a distributed load-balanced routing scheme based on Software Defined Networking. The approach divides the large-scale network into sub-sections, called clusters. In order to minimize signaling overhead, packets are forwarded between these clusters according to geographical heuristics. Within each cluster active Quality of Service-aware load-balancing is applied. The responsible on-board network controller forwards routing instructions based on the network state information in its cluster. We also analyze specific design choices for the clusters and the interfaces between them. The protocol has been implemented in a system-level simulator and compared to a source-routed benchmark solution. 
\end{abstract}


\section{Introduction}
\label{sec:introduction}
In recent years, there has been increased interest into non-terrestrial networks capable of providing global broadband access. Large Low Earth Orbit (LEO) Satellite Constellation Networks (SCNs) in particular are gaining more attention, most notably because of their proximity to Earth, which results in reduced propagation delays in comparison to \ac{GEO} satellites. Prominent examples for such constellations using circular orbits include OneWeb, and Starlink by SpaceX, among others \cite{butash_nongeostationarysatelliteorbit_2021}. Moreover, current initiatives for future telecommunication standards, such as 5G \cite{3gpp_solutions_2019}, consider the integration of such \ac{LEO} \acp{SCN}. In order to reduce the required number of \acp{GW} on ground, most designs consider \acp{ISL}. In this work, we assume that broadband multimedia voice and data traffic, all based upon IP, will be routed via a single layer \ac{SCN}. The connectivity for private end users is provided via an IP interface attached to a \ac{UT} on ground. \acp{GW} act as interfaces to the terrestrial Internet. The considered \ac{SCN} needs to provide bidirectional connectivity at IP layer. Internally, the \ac{SCN} may use other protocols (e.g. on layer 2) for routing and switching.

The SCN network topology differs from terrestrial networks in multiple aspects. Firstly, the network topology is highly dynamic, but deterministic. The connectivity between satellites changes depending on the constellation design pattern. These changes in the \ac{ISL} connectivity are typically predictable. In addition, there are frequent handover events with the terminals on the ground. Due to the movement speed of the satellites relative to the Earth in \ac{LEO}, their visibility is limited. Therefore, a terminal on ground has to establish a connection to a new satellite in order to maintain its connection to the SCN. It is only possible to predict to which satellite a new \acp{ESL} is established, if a stringent handover strategy is used. However, we intend to include handover schemes which also consider the current load on the satellites in the future. Thus, we assume that these ESL handovers are not fully predictable. 

In general, the location of \acp{UT} on ground depends on geographical and demographic factors. Therefore, resulting traffic is non-uniformly distributed. Traffic hot-spots are geographical rather than topological, due to the relative movement of the space segment.
As we consider large constellations, communicating current link states to all relevant nodes is very costly in terms of signaling overhead. Due to the physical size of the network, the signaling also experiences significant propagation delays.
In addition, processing resources on-board the satellites are scarce. Thus, \ac{LEO} \acp{SCN} with \acp{ISL} need specialized routing and network management protocols to make the best use of the available network resources. While the topological changes of the constellations are highly predictable, the traffic demands are not. Therefore, we focus on routing protocols with load-balancing capabilities.

Based on these considerations and related research we propose a distributed routing scheme which proactively balances the network load. The approach utilizes distributed on-board \ac{SDN} controllers, which enable load-balanced routing and network management within their respective clusters. The routing between clusters follows geographic heuristics in order to minimize signaling overhead. The main contributions of this paper include:
\begin{itemize}
    \item A distributed load-balancing routing scheme based on SDN is proposed. It was designed to reduce signaling overhead, while providing locally centralized congestion control and low end-to-end latency.
    \item Relevant specific design choices for a realistic system setup for polar LEO SCNs are discussed.
    \item The proposed scheme is implemented in a system-level simulator and compared to a source-based routing protocol. The specific advantages of the proposed protocol are illustrated.
\end{itemize}

The rest of the paper is organized as follows. In Section \ref{sec:related}, we shortly cover related works. Then, in Section \ref{sec:system} the considered reference system is presented. The proposed routing protocol is described in detail in Section \ref{sec:routing}. In Section \ref{sec:results}, we provide simulation results and comparisons to benchmark solutions. Lastly, Section \ref{sec:conclusion} concludes the paper.

\section{Related Works}
\label{sec:related}
Different routing strategies have been proposed in the literature for LEO SCNs. On the one hand, there are decentralized approaches, such as \ac{ELB} \cite{taleb_sat043elbexplicit_2006}. Such schemes are based on locally transmitted signals which alarm neighboring nodes of congested links or buffers.
As such, they enable local load-balancing with very little signaling overhead.
In the case of ELB, pre-determined thresholds were utilized, which triggered the signaling of a “fairly busy” and a “busy” state. The downside of completely decentralized approaches is their limited scope. They are unable to identify and detect hot spots proactively and distribute the network load accordingly. This can result in traffic flow cascade problems. The overloading of a single node results in a local rerouting to neighboring nodes - which in turn causes additional overloaded nodes nearby. Emerging hot spot cannot be circumvented directly. As the overloaded state propagates, unexpected packet detours or even loops are possible depending on the protocol design.

On the other hand, dedicated load-balancing approaches with a larger scope may make better use of the available network resources. Terrestrial load-balancing schemes, e.g. B4 \cite{jain_b4experiencegloballydeployed_2013}, enable proactive routing with low latency and are less prone to congestion. Yet, these approaches do not update as frequently as required by the dynamic topology changes of SCNs. Still, the effectiveness of such schemes when tailored specifically to such networks is relatively unexplored. Naturally, the monitoring of multiple link states results in additional signaling overhead. Therefore, we are primarily interested in protocols which minimize this overhead.

The required signaling can also be significantly reduced if the handovers between satellites and terminals on ground are resolved locally. Then, network-wide mapping updates are not required. This can be achieved by utilizing fixed virtual or logical nodes.
A particularly useful approach is geographical routing, which utilizes geographical areas as virtual nodes. Terminals on ground are given an area identifier which is used by the SCN to forward the packets in the direction of the destination region \cite{henderson_distributed_2000, tsunoda_supportingipleo_2004}. Additional mechanisms are required to circumvent constellation-specific constraints efficiently and solve ambiguities in the last hop \cite{tsunoda_geographicalorbitalinformation_2006}. Mobile terminals also present a challenge to this concept, as they change their geographical position, which needs to be identified and communicated in the SCN.
In our recent work \cite{roth_implementationgeographicalrouting_2021}, geographical identifiers were placed in mutable layer 2 addresses as opposed to the typical placement in layer 3 addresses. This allowed for a decoupling between the address resolution and the routing. Updates were only required when new users joined the network, or when a mobile user entered a different area and was assigned a new identifier. The approach offered an attractive trade-off between satellite autonomy, latency and reduced signaling. However, only carefully designed local load-balancing and rerouting mechanisms were possible which can be an issue for broadband traffic.

To enable proactive load-balancing in a larger domain, some degree of centralization is required. In this context, using \ac{SDN} can enable efficient decision-making. A key advantages of SDN-based protocols is their flexibility via softwarization and virtualization. The programmability of \acp{SDN} can provide fast reconfiguration, increased resiliency, service-aware \ac{QoS}, and dynamic network topology support \cite{bertaux_softwaredefinednetworking_2015}. But, complete centralization is impractical for SCNS due to their physical size, which results in significant propagation delays, and the required signalling. A more promising approach for SCNs is distributed SDN, which enables centralized control in a smaller scope \cite{papa_dynamicsdncontroller_2018, papa_designevaluationreconfigurable_2020}.
A flat control architecture with on-board SDN controllers was used in corresponding investigations.
An \ac{SPF} algorithm was used to realize the routing decisions.
We consider such frameworks particularly advantageous for SCNs, as they provide useful trade-offs. It can combine advantages of decentralized schemes, namely reduced signaling and fast reactivity, as well as advantages from centralized approaches, such as proactive network-control. Therefore, the proposed protocol is based on a distributed SDN with a flat control hierarchy. The disadvantages of such schemes is the more involved cluster design and the handling of interfaces between them.

\section{Reference System}
\label{sec:system}
\subsection{Space segment}
For this investigation, we will investigate a Walker star mega-constellation \cite{walker_satelliteconstellations_1984}. This design pattern is characterized by polar or near-polar orbits.
This results in adjacent counter-rotating planes, which are called seams. Due to the high Doppler between these planes, we assume that no ISL connection is possible. This constellation-specific constraint has to be respected by geographical routing approaches. The design is based on the Iridium constellation \cite{fossa_overviewiridiumlow_1998}. Recent projects, such as OneWeb, deploy similar Walker star constellations. This design pattern provides global coverage, and typically faster north-to-south, and vice versa, connections than inclined constellations \cite{handley_delaynotoption_2018}.
On the other hand, latitudinal routes are less efficient. So, this design corresponds less to the geographical demand distribution over this axis.

We assume that each satellite has four \acp{ISL}: two intra-plane \acp{ISL} and two inter-plane \acp{ISL}. The spacing of the co-rotating and counter-rotating planes follows the standard design pattern for Walker star constellations. We assume $1440$ satellites in $30$ planes, with $48$ satellites per plane. Based on the near-polar orbits and the number of satellites, the reference constellation is called P-1440. In addition, a shutdown of the inter-plane ISLs is required at the polar regions, because of the switch in relative position of the orbital planes. 
Due to the relative speed and angle of satellites, inter-plane ISLs are typically deactivated to readjust.
The shutdown latitudes mainly depend on the pivoting speed of the antennas and the angular velocity between the two satellites.
As we assume state-of-the-art optical laser terminals are used for the \acp{ISL}, an inter-plane shutdown latitude of $\pm80^\circ$ is considered feasible. The characteristics of the constellation are summarized in Tab. \ref{tab:reference}. We assume an ISL data rate of $1$Gbps.

\begin{table}
\caption{Summary of P-1440 constellation and ground segment.}
\label{tab:reference}
\def\arraystretch{1.3}
\begin{center}
\begin{tabular}{|l|c|}
    \hline
    \textbf{Space segment characteristic}& 
    \textbf{Value}\\
    \hline
    Number of satellites&
    1440\\
    \hline
    Number of planes& 
    30\\
    \hline
    Number of satellites per plane& 
    48\\
    \hline
    Satellite altitude [km]& 
    600\\
    \hline
    Orbital inclination [$^\circ$]& 
    86.4\\
    \hline
    Cross-seam planes spacing [$^\circ$]& 
    12\\
    \hline
    Co-rotating planes spacing [$^\circ$]& 
    6\\
    \hline
    Angular phase offset co-rotating planes [$^\circ$]& 
    3.75\\
    \hline
    Inter-plane ISL shut-down latitude [$^\circ$]& 
    80\\
    \hline
    \hline
    \textbf{Ground segment characteristic}& 
    \textbf{Value}\\
    \hline
    \hline
    Number of user terminals& 
    10000\\
    \hline
    User terminal minimum elevation angle [$^\circ$]& 
    30\\
    \hline
    Number of gateway stations& 
    39\\
    \hline
    Gateway minimum elevation angle [$^\circ$]& 
    20\\
    \hline
\end{tabular}
\end{center}
\end{table}

\subsection{Ground segment}
The ground segment of our reference system consists of \acp{UT} and \acp{GW}. \acp{UT} represent private or business user terminals. The geographic distribution of the \acp{UT} is based on global population density statistics \cite{centerforinternationalearthscienceinformationnetwork-ciesin-columbiauniversity_griddedpopulationworld_2018}. However, metropolitan and densely populated areas need to be discounted, as they tend to be well connected. So, we applied a scaling function which adjusts high density values. The result is then fed into the stochastic process to generate a relevant user terminal distribution. The actual population density is used until a density of $100$ persons per square kilometer is reached. From this point on, all areas are limited to an adjusted density of $100$. 
Based on this adjusted population density, the terminal distribution shown in Fig. \ref{fig:distribution} has been generated \cite{cartopy}. To this end, the two-dimensional density distribution has been sampled using the corresponding cumulative density functions. $10 000$ end user terminals were generated for the figure.

For the minimum elevation angle of the user terminals, various aspects of the reference constellation have to be taken into consideration. In general, higher elevation angles allow for higher available data rates in the satellite communication channel, e.g. by using Adaptive Coding \& Modulation schemes. However, this results in a higher number of handover events. The minimum elevation angle also determines the number of simultaneously visible satellites. For P-1440, we consider a minimum of four satellites to be sufficiently diverse, and therefore assume a minimum elevation angle of $30^\circ$. The corresponding maximum visibility of a satellite fly-by is $236$s.

The locations of the GWs are chosen more deliberately.
In theory, an optimal GW distribution can be computed using the expected UT density with minimal overlap between GWs. But in reality, they are placed strategically with regards to existing Internet access infrastructure. Therefore, we preferred locations near cities with data centers. In addition, we considered the physical distance between the UTs and the closest GW. Since some UTs are on islands or in remote regions, we chose not to decide on a minimum physical distance for all UTs. The resulting GW distribution is non-optimal, but near cities with existing infrastructure.
Based on the overall expected traffic volume, we assume 39 gateway campuses for our scenario. As dedicated structures, we consider GWs to achieve a minimum elevation of $20^\circ$. Their distribution is shown in Fig. \ref{fig:distribution} as well. The characteristics of the ground segment are summarized in Tab. \ref{tab:reference}. The impact of the GW distribution depends on the traffic models. In this reference system, we assume that users connect to all GWs. We consider connections to far-away GWs, as the end-to-end latency is lower compared to terrestrial networks (although content is oftentimes cached nearby). If we were to focus solely on the closest GW as potential destinations, the latency would depend more on their distribution.

\begin{figure*}[htbp]
\centering{\includegraphics[width=0.9\textwidth]{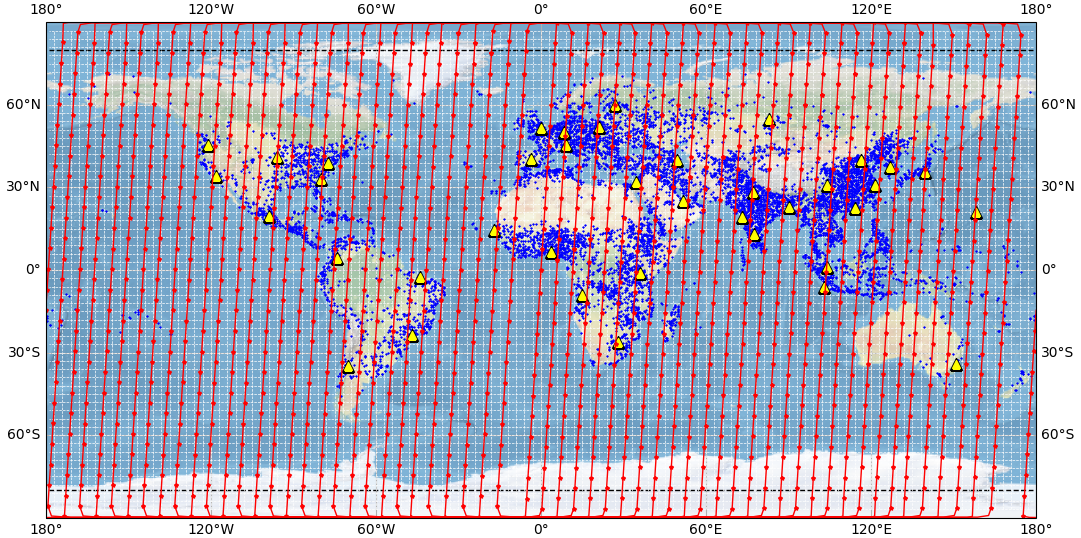}}
\caption{Reference system: P-1440 constellation (red), UTs (blue) and GWs (yellow), inter-plane ISL shutdown (black), geographical areas (white).}
\label{fig:distribution}
\end{figure*}

\subsection{Connectivity}
\label{sec:traffic}
For both ground segment types, seamless handovers are assumed. In this context, a seamless handover describes a fast switch which is not perceptible by a user. However, a \ac{UT} is not necessarily connected to two satellites at the same time. We assume that \acp{UT} use phased array antenna technology, which allows for fast and split pivoting of beams.
So, we assume that a simultaneous connection to two satellites is possible.
Handovers can be prepared beforehand, it is a make-before-break strategy. In terms of connectivity, it is generally assumed that a \ac{UT} is only sending and receiving data via a single \ac{ESL}. We assume a maximum \ac{ESL} downlink data rate of $1$Gbps. The aggregated uplink ESL data rate is also $1$Gbps. An individual UT is assumed to establish a session with an average data rate of $20$Mbps. As we group UTs in the simulation, we assume an aggregated data rate of $100$Mbps. 
In this initial investigation, a simplified traffic model is assumed: there are no \ac{QoS} classes with varying priorities or delay budgets. The assumed data rates and traffic characteristics are also summarized in Tab. \ref{tab:sim}.

\section{Routing Approach}
\label{sec:routing}
The solution we propose is named \ac{IDLB}. The approach aims to reduce signaling overhead while using network resources efficiently. To this end, SDN clusters exchange minimal network information between each other. Each cluster calculates routes independently of the network state of its neighbouring clusters. Only the satellites of other clusters at the interfaces are monitored. This is necessary to avoid forwarding packets into hot spots at the interfaces. The intra-cluster routing uses proactive load-balancing. The inter-cluster routing is based on geographical heuristics to minimize signaling overhead. The main drawback of this approach is that the resulting routes are potentially less efficient in comparison to end-to-end routing approaches. Due to the limited scope, traversing the interfaces between clusters can be sub-optimal.

The scheme is based on space-borne distributed SDN with a flat control hierarchy. We consider the positions of the clusters fixed within the constellation, the SDN controllers are static. In contrast to fixed clusters based on geographical locations, this approach does not require logical handovers due to the movement of the satellites. Frequent migrations of the role of the SDN controller as well as new flow instructions for nodes entering a different cluster can also be avoided. The mapping between the in-orbit topology and on-ground geography can be computed without any signaling.

The proposed solution is distinct from previous protocols, because of the combination of load-balancing broadband traffic with distributed SDN in a SCN context. To the best of our knowledge, there is no current approach which uses dedicated load-balancing for intra-cluster routing and geographical heuristics for inter-cluster routing. The approach is expected to provide minimal packet dropping rates and signaling overhead, while offering similar latencies as exhaustive \ac{SPF}-based approaches.
While the proposed protocol was conceptualized for Walker Star constellations, adjustments for other constellation types may be investigated in future work.

\subsection{Geographical address resolution}
A workable routing protocol also requires an explicit address resolution scheme suitable for a real-world system. As the satellites are in constant movement, there are frequent handover events with terminals on ground. Thus, it is difficult to keep track of the current \ac{ESL} assignment. It is not obvious which satellite is currently connected to the requested UT or GW, unless there are stringent handover rules. If we allow for more flexible and adaptive handover strategies, ambiguities in the last hop cannot be avoided. Thus, signaling is required to communicate a handover event to all interested entities.

We consider a geographical aggregation of addresses \cite{maihofer_surveygeocastrouting_2004}. Such a geographical address resolution scheme typically uses a geographical marker in a terminal-specific identifier. Traffic can be routed towards the geographical area of the destination until the address can be resolved locally. The identifier can be placed in the layer 2 \cite{roth_implementationgeographicalrouting_2021} or the layer 3 address \cite{tsunoda_geographicalorbitalinformation_2006} of a packet.

The size of the geographical areas is relevant. Smaller areas result in less ambiguity, as the set of potential serving satellites is smaller. On the other hand, markers are longer and more signaling is caused by mobile terminals. For the reference scenario, we propose areas of $3^\circ$ latitude and $3^\circ$ longitude up to $\pm87^\circ$. At the poles, single cells are used. This results in $6962$ areas, which can be represented by $13$bits.

\subsection{Cluster design}
The physical size of a cluster is crucial for the efficiency of the protocol. A larger scope is useful for more efficient load-balancing. However, the latency between the SDN controller and its forwarding devices increases. When we approach a centralized system, the reactivity decreases, while convergence time and signaling overhead increase. Thus, to find an appealing trade-off, the upper limit of the cluster size should be determined by the longest latency within the cluster between the controller and a corresponding node. We propose that this maximum intra-cluster latency should be less than $50$ms. With a \ac{RTT} of $100$ms, an adequate responsiveness to failures, unexpected events or traffic is expected.

For P-1440, we consider a rectangular arrangement of $6\times8$ nodes (a cluster size of $48$) to be a good initial approach. This results in $30$ clusters, each cluster representing $3.33\%$ of the satellites of the constellation. The expected maximum intra-cluster propagation delay is $39.7$ms. While the average number of hops and intra-cluster latency can be decreased with different cluster arrangements, they do not provide clean transitions at the seams. We focus on rectangular arrangement, because all clusters possess the same shape and number of satellites - there are no cropped clusters at the seams.

\subsection{Load-balanced intra-cluster routing}
\label{sec:protocol}
The intra-cluster routing of IDLB relies on efficient routes and proactive load-balancing. We try to avoid the creation of bottlenecks or to circumvent them if required. Since packet drops are costly in an SCN context due to high propagation delays, load-balancing strategies are fundamental for supporting broadband traffic requirements. As we operate on a tightly meshed network, spreading the traffic does not necessarily result in significantly increased latencies. There are typically multiple routes of similar propagation delay \cite{handley_delaynotoption_2018}, which is the principal factor for the end-to-end latency. For instance, a QoS-based distribution of the traffic can improve the latency of delay-critical classes: the shortest paths are not blocked by established flows with lower priority.

In order to establish proactive load-balancing, the SDN controller has to apply a suitable path-selection strategy. From a theoretical point of view, a max-flow min-cut-based route selection algorithm \cite{alzaben_endtoendroutingsdn_2021} is a useful strategy. By determining the minimum cut between a node within the cluster and an interface node, we get the set of edges which represent the main bottleneck. Then, we can choose the least congested edge which still results in a path with the minimum number of hops. Overloaded links are circumvented to avoid drops. Another promising path selection approach is a k-shortest paths approach \cite{eppstein_findingshortestpaths_1997}. This enables per-packet or per-flow QoS-based multipath routing within the cluster. In this initial investigation, we use a dynamic shortest path approach, which circumvents links with high loads. If the preferred interface is overloaded, a different path is used to exit the cluster. 

The routes are computed periodically by the SDN controller based on the network state information within its cluster. The individual flow instructions are forwarded to the nodes of the cluster. Changes in the network topology can also trigger the distribution of new routes. Similarly, a cluster will also update its mappings when handing over a terminal in an active session to another cluster. By using such proactive updates, the proposed protocol was free of deadlocks and loops in our simulations.

\begin{figure}[htbp]
\centering{\includegraphics[width=0.5\textwidth]{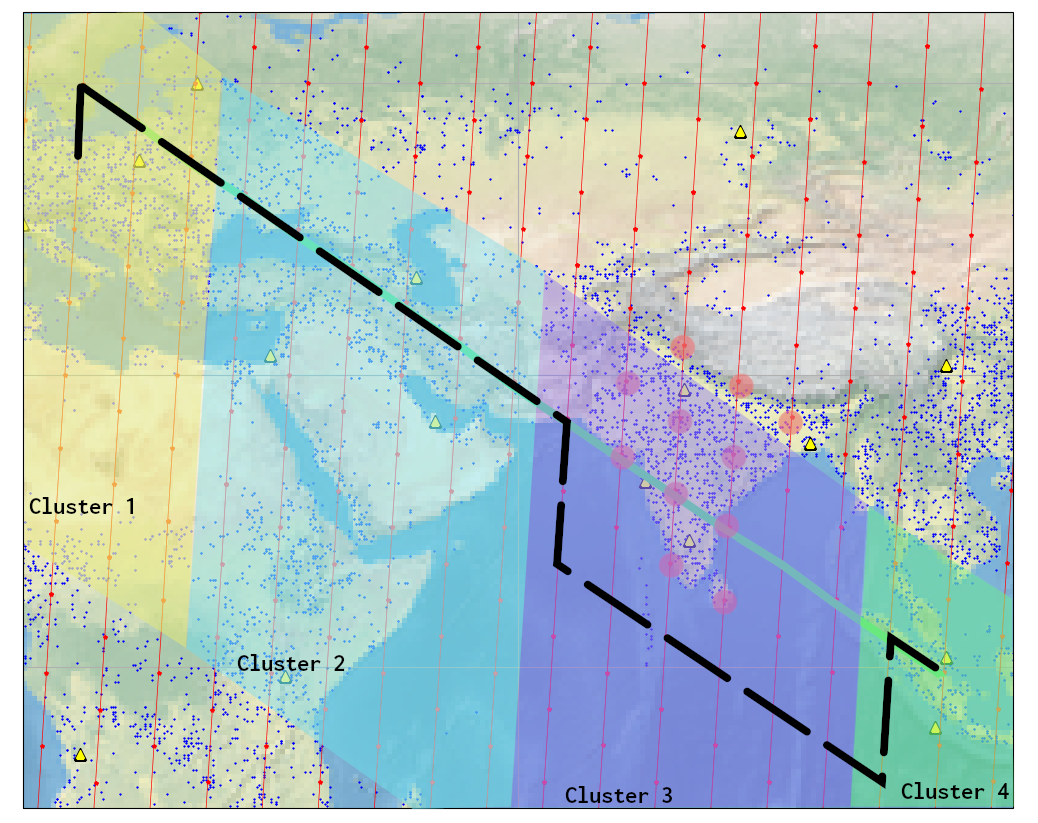}}
\caption{Exemplary route with overloaded nodes (red) in a cluster.}
\label{fig:path}
\end{figure}

\subsection{Geographic-based inter-cluster routing}
Between clusters, packets are routed based on the geographic information of the destination terminal. So, if the serving satellite is not within the current cluster, the packets are forwarded towards a cluster which is in proximity of the geographical area. As mentioned, forwarding based on a geographical identifier can be ambiguous.

This can not be easily resolved by forwarding to a neighboring cluster if we are at a seam. So, the geographical heuristic has to respect that no transmission across the seam is possible. Otherwise, the seam may be approached from the wrong side, which results in multiple additional hops. We achieve this via an internal network representation based on the clusters.
In addition, all UTs and GWs of a geographical cell at the seam have to handover at a distinct moment in time. This way, the switch over to the other side of the seam can be handled proactively by every SDN controller with minimal signalling.

The direction then determines the set of potential interface nodes. Since the intra-cluster routing computes routes to all interface nodes of neighboring clusters, no additional path selection is required.

In general, there is a trade-off between end-to-end route optimization and signaling overhead. With IDLB, we are only aware of network state in a limited scope, the resulting path may still be sub-optimal. As the next cluster calculates their intra-cluster paths independently, zig-zagging end-to-end paths may arise due to the individual load-balancing.
However, a quantified comparison is required to determine the impact of sub-optimal interfaces, because of the densely meshed structure of SCNs.
By exchanging network state information with neighboring clusters, forwarding at the interfaces can be improved. As this approach results in additional signaling overhead, it is not considered in this investigation.
An exemplary route traversing four clusters is shown in Fig. \ref{fig:path}. Packets are rerouted in in cluster 3, due to overloaded nodes.

\begin{figure}[!t]
\removelatexerror
\begin{algorithm}[H]
    \caption{Network update for IDLB (no priority)}
    \SetKwInOut{Input}{Input}
    \SetKwInOut{Output}{Output}
    \Input{connectivity, network state information}
    \Output{routes, update instructions}
    \For(check status){$n_{(i)}$ in $n_{c}$ and all $n_{(i)}$ in $n_{int}$}
    {
        \uIf{no update}
        {
            delete from $active\_n_{c}$ or $active\_n_{int}$\;
            update topology\;
            set failure flag for $n_{(i)}$\;
        }
        \ElseIf{overload $\algorithmicor$ inter-plane warning}
        {
            update topology\;
            set flag for $n_{(i)}$\;
        }
    }
    \For(update inter-cluster) {$n_{(i)}$ in $active\_n_{int}$}
    {
        \For(set nearest cluster) {$area_{(i)}$ in $areas$}
        {
            update geographic switching table;
        }
    }
    \For(update intra-cluster){$n_{(i)}$ in $active\_n_{c}$}
    {
        \uIf{(flow-specific) interface adjustment}
        {
            $n_{dst} := n_{proposed}$\;
        }
        \Else
        {
            set of $n_{dst}$ from geographic switching table\;
        }
        find weighted k-shortest paths $n_{src} \rightarrow n_{dst}$\;
        set instructions\;
        distribute\;
    }
\end{algorithm}
\end{figure}

\subsection{Signaling overhead}
\label{sec:signaling}
The management cost $C$ can be used to evaluate the signaling overhead. With the message size $M$ and the required number of hops $H$ the cost can be defined as \cite{roth_implementationgeographicalrouting_2021}: $C = M \cdot H$. We also include the update frequency of $R^{update}$, to calculate the time-dependent cost $C(t)$.
With distributed SDN, the periodical updates come from the forwarding devices in a cluster ($n_{c}$) to the responsible SDN controller and vice versa. The average number of hops within a cluster $H_{c}$ depends on the distance of each satellite to the SDN controller. We consider a transmission of network state information from the nodes, and flow instructions to the nodes (thus a factor of $2$). In addition, we assume an exchange with bordering interface nodes from neighboring clusters is necessary. The number of such interface nodes is $n_{int}$, their average number of hops to the SDN controller is $H_{int}$. We can describe the expected signaling overhead in a cluster as:
$$C_{c}(t) \approx M \cdot (2\cdot H_{c}(n_{c}-1) + H_{int}n_{int}) \cdot R_{c}^{update}$$

As the IDLB protocols minimizes the overhead between clusters, no routing-related signaling is necessarily required. With the proposed cluster size and arrangement, we have $30$ individual, identical clusters. But, clusters at the seams have fewer interface nodes. Thus, the overall management cost of IDLB can be described by:
$$C_{IDLB}(t) \approx N_{c; seamless} \cdot C_{c; seamless} + N_{c; seam} \cdot C_{c; seam}$$

With the assumed rectangular cluster design, the SDN controller is near but not directly in the middle of the cluster. By summing the minimum number of hops required to reach the SDN controller for each node in a cluster, we can calculate the average number of hops for P-1440: $H_{c} \approx 3.57$. The complete management cost for IDLB is approximately $14304$ packets per network update. For this cluster design, requesting network state information from neighboring clusters results in a signaling increase of at least $14\%$.

\subsection{Latency}
\label{sec:latency}
The end-to-end latency is a key performance indicator for a routing protocol. We denote $t_r^{\left(i\right)}$ as the time required for the packet routing decision at node $(i)$. We use $t_s^{\left(i\right)}$ the time spent for switching on the node. Both of these quantities are expected to be in the domain of $\mu s$. $W_{input\rightarrow output}^{\left(i\right)}$ is used to describe be the queuing delay on a node. Finally, we denote  $D_{prop}^{\left(i\right)\rightarrow\left(i+1\right)}$ the propagation delay between nodes $(i)$ and $(i+1)$. We thus propose the following description of the end-to-end latency of a packet in an SCN:
$$L^{src\rightarrow dst}=\sum_{i:src}^{dst} t_r^{\left(i\right)} +t_s^{\left(i\right)} +W_{input\rightarrow output}^{\left(i\right)} +D_{prop}^{\left(i\right)\rightarrow\left(i+1\right)}$$

The queuing delay mainly depends on the size of the on-board buffers and the efficiency of load-balancing. In general, we want to keep queues rather small in order to reduce the maximum delay on a hop. Typically, for burst-like broadband traffic, queues are either empty or fill up quickly.  
So, larger queues do not necessarily provide an added benefit. Typically, the propagation delay is the most significant quantity impacting the latency. For the proposed P-1440 constellation, the ISL propagation delay between nodes is approximately between $[3$ms, $5.8$ms$]$. However, for scenarios with elevated traffic and large buffer of packet sizes, similar queuing delay are possible. Thus, the routing scheme has to provide low propagation delays and use the available network resources efficiently. Due to the tightly-meshed network, routes with an equal number of hops can result in similar end-to-end latency. Therefore, efficient load-balancing is oftentimes possible while maintaining close-to-optimal latency, which is a design goal of the proposed protocol.

\section{Performance Evaluation}
\label{sec:results}

\subsection{Simulation setup}
The described reference system was implemented in a system-level software simulator. The simulator was written in C++ as an extension of the simulator described in \cite{roth_implementationgeographicalrouting_2021}. The simulation parameters are summarized in Tab. \ref{tab:sim}.

For the simulations, we assumed that $2000$ randomly chosen UTs are active, and that they have an aggregated uplink feeder data rate of $100$Mbps. All users are are served fairly, there is no inherent preference or priority. The traffic consists of individual sessions between randomly chosen terminals on ground.
The data rate of each session is constant.
$50\%$ of the sessions are between GWs and UTs, the other $50\%$ are UT-to-UT. The starting times of the sessions are distributed uniformly throughout the simulation. Their duration follows a Poisson distribution. The packet and buffer sizes have been scaled to accelerate the simulations. The maximum queuing delay of a single hop is approximately $6$ms, which is in the same order of magnitude as the propagation delays. For the given simulation duration, the satellites orbit the Earth completely. Therefore, the simulations include a high number of topological changes due to handover events and inter-plane ISL shutdowns.
The correctness of the routing at the seams is demonstrated by the simulation, as the seam moves over Europe, a region with a high UT and GW density.

To evaluate the efficiency of the proposed IDLB scheme, it is compared to a source-routed \ac{SPF} approach.
This scheme represents a realistically applicable routing solution, which provides fast routes with minimal signalling. As we only consider a single QoS priority, the approach is particularly useful to evaluate the impact of the load-balancing w.r.t. latency.
The source-routed approach relies on an \ac{SPF} algorithm to compute the shortest path between two nodes without considering individual link loads. We thus expect significantly higher dropping rates.
The implementations of both schemes circumvent the inter-plane ISLs of nodes, which are about to enter the inter-plane shutdown latitudes. Therefore, there should be no drops due to this temporary deactivation.

\begin{table}
\caption{Summary of simulation parameters.}
\label{tab:sim}
\def\arraystretch{1.3}
\begin{center}
\begin{tabular}{|l|c|}
    \hline
    \textbf{Simulation parameter}& 
    \textbf{Value}\\
    \hline
    \hline
    Simulation duration [s]& 
    7200\\
    \hline
    Precision of movement [s]& 
    5\\
    \hline
    \hline
    ISL data rate [Mbps]& 
    1000\\
    \hline
    On-board output buffer size [Mbit]& 
    6\\
    \hline
    Aggregated maximum feeder downlink [Mbps]& 
    1000\\
    \hline
    Aggregated maximum feeder uplink [Mbps]& 
    5000\\
    \hline
    \hline
    Number of simultaneously active UTs&
    2000\\
    \hline
    Number of sessions& 
    10 000 -- 20 000\\
    \hline
    Average session duration [s]& 
    30\\
    \hline
    Aggregated data rate per user terminal [Mbps]&
    100\\
    \hline
    Packet size [Mbit]&
    1.2\\
    \hline
    Maximum number of hops&
    60\\
    \hline
\end{tabular}
\end{center}
\end{table}

\subsection{Results and analysis}
We focus on the average end-to-end delay of a session, and the dropping rate. The signaling overhead has been formulated and described in Sec. \ref{sec:signaling}. To reiterate, the proposed scheme offers significantly less signaling overhead than comparable source-routed approaches. The signaling overhead is also lower compared to distributed schemes which share additional information between clusters.

A comparison of the end-to-end latency for $20 000$ sessions is shown in Fig. \ref{fig:latency}. Because of the demanding network load, we expect longer end-to-end latencies due to queuing delays. On average, the session delays between this initial implementation of IDLB and the SPF-based approach are comparable. The intersection value of the histograms is $0.65$. The result is expected for load-balancing solutions, as bottlenecks of the network are actively circumvented. In addition, we compute end-to-end routes only within clusters, which can also contribute to potentially longer paths.

\begin{figure}[htbp]
\centering{\includegraphics[width=0.5\textwidth]{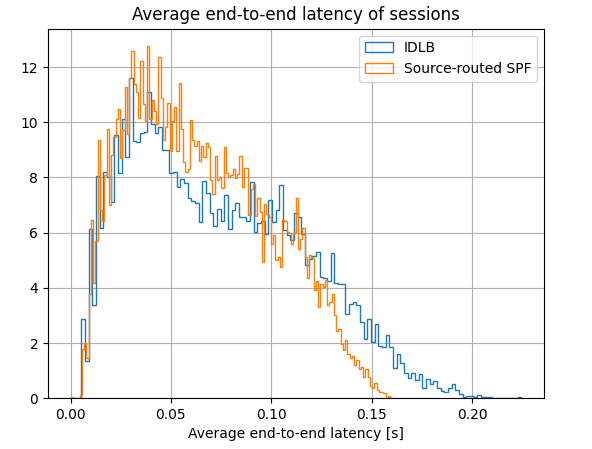}}
\caption{Comparison of end-to-end latency for 20 000 sessions.}
\label{fig:latency}
\end{figure}

Fig. \ref{fig:drops} shows a comparison of the packet dropping rates for different network loads. As expected, IDLB causes significantly fewer drops compared to the SPF-based approach. In the traffic scenario with the highest network load, consisting of $20 000$ sessions, the dropping rate of IDLB is $3.6\cdot10^{-4}$. For the same network load, the source-routed benchmark drops approximately $1\%$ ($10^{-2}$) of the packets. Using Little's law \cite{little2008little}, we can estimate that the average number of simultaneously active sessions is approximately $83.3$. Since a session has a data rate of $100$Mbps, $10$ simultaneous sessions on a single link use all of the capacity ($1$Gbps). Given the non-uniform distribution of terminals, active load-balancing is thus required to mitigate packet drops. By circumventing hot spots, the load on specific links and nodes was lightened with IDLB.

For the shown lower traffic loads ($10 000$ and $15 000$ sessions), IDLB also outperforms the source-routed approach. Naturally, if there are few congested links or nodes, the difference in dropping rates is minimal. The downlink and uplink limitations of the satellites also cause drops if the satellite is connected to frequently requested terminals on ground. Thus, handover strategies which also consider the current load on a node could further decrease the dropping rates. Nevertheless, in the given scenarios, the amount of packets dropped due to downlink limitations was comparatively low. For the most demanding traffic scenario ($20 000$ sessions), the dropping rate due to overflows at the downlink buffers was only $7\cdot 10^{-5}$.
With a lower number of active UTs (1000), more of the traffic was concentrated between high-density regions. Nevertheless, the number of drops only varied slightly indicating that overlap of routes did not increase significantly. Defining different UT and GW distributions should impact the results more, but is not in the scope of this work.

In the simulations, we assumed constant data rates for all sessions with equal priority. However, Internet traffic is typically bursty, with multiple priorities. In such cases, we expect bottlenecks to appear more sudden and to be more transient. Depending on the frequency of network updates within a cluster, this may result in more packet drops. The impact of more realistic traffic types will be analyzed in future investigations.

\begin{figure}[htbp]
\centering{\includegraphics[width=0.5\textwidth]{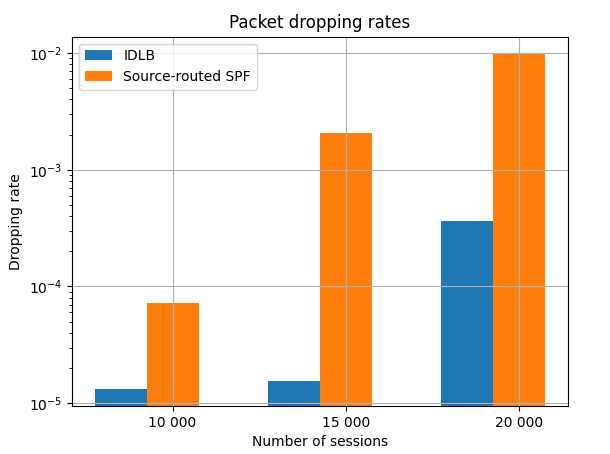}}
\caption{Comparison of packet dropping rates for different traffic loads.}
\label{fig:drops}
\end{figure}

\section{Conclusion}
\label{sec:conclusion}
In this work, we proposed an initial design of the IDLB protocol, a distributed load-balancing protocol specifically designed to handle broadband traffic in LEO mega-constellations with ISLs. The proposed scheme is based on clusters with individual SDN controllers on board dedicated satellites. Within a cluster, routes are computed based on proactive load-balancing. The routing between clusters is determined by geographical heuristics and signaling at the interfaces. Corresponding design choices, such as the geographical addressing scheme and the cluster design were discussed as well. We compared the proposed protocol to a source-routed SPF strategy. The results illustrated that a comparable end-to-end latency can be achieved with drastically decreased dropping rates and significantly lower signaling overhead. Further in-depth analyses with multiple \ac{QoS} profiles, varied traffic scenarios, and more extensive simulations are planned.

\bibliographystyle{IEEEtran}
\bibliography{IEEEabrv, distributed-load-balanced-routing}

\begin{thebibliography}{10}
\providecommand{\url}[1]{#1}
\csname url@samestyle\endcsname
\providecommand{\newblock}{\relax}
\providecommand{\bibinfo}[2]{#2}
\providecommand{\BIBentrySTDinterwordspacing}{\spaceskip=0pt\relax}
\providecommand{\BIBentryALTinterwordstretchfactor}{4}
\providecommand{\BIBentryALTinterwordspacing}{\spaceskip=\fontdimen2\font plus
\BIBentryALTinterwordstretchfactor\fontdimen3\font minus
  \fontdimen4\font\relax}
\providecommand{\BIBforeignlanguage}[2]{{%
\expandafter\ifx\csname l@#1\endcsname\relax
\typeout{** WARNING: IEEEtran.bst: No hyphenation pattern has been}%
\typeout{** loaded for the language `#1'. Using the pattern for}%
\typeout{** the default language instead.}%
\else
\language=\csname l@#1\endcsname
\fi
#2}}
\providecommand{\BIBdecl}{\relax}
\BIBdecl

\bibitem{butash_nongeostationarysatelliteorbit_2021}
T.~Butash, P.~Garland, and B.~Evans, ``Non-geostationary satellite orbit
  communications satellite constellations history,'' \emph{International
  Journal of Satellite Communications and Networking}, vol.~39, no.~1, pp.
  1--5, 2021.

\bibitem{3gpp_solutions_2019}
{3GPP}, ``Solutions for {{NR}} to support {{Non-Terrestrial Networks}}
  ({{NTN}}),'' {3GPP}, Tech. Rep. TR 38.821 V16.0.0, Dec. 2019.

\bibitem{taleb_sat043elbexplicit_2006}
T.~Taleb, D.~Mashimo, A.~Jamalipour, K.~Hashimoto, Y.~Nemoto, and N.~Kato,
  ``{{SAT04-3}}: {{ELB}}: {{An Explicit Load Balancing Routing Protocol}} for
  {{Multi-Hop NGEO Satellite Constellations}},'' in \emph{{{IEEE Globecom}}
  2006}, Nov. 2006, pp. 1--5.

\bibitem{jain_b4experiencegloballydeployed_2013}
S.~Jain, A.~Kumar, S.~Mandal, J.~Ong, L.~Poutievski, A.~Singh, S.~Venkata,
  J.~Wanderer, J.~Zhou, M.~Zhu, J.~Zolla, U.~H{\"o}lzle, S.~Stuart, and
  A.~Vahdat, ``B4: Experience with a globally-deployed software defined wan,''
  \emph{ACM SIGCOMM Computer Communication Review}, vol.~43, no.~4, pp. 3--14,
  Aug. 2013.

\bibitem{henderson_distributed_2000}
T.~Henderson and R.~Katz, ``On distributed, geographic-based packet routing for
  {{LEO}} satellite networks,'' in \emph{Globecom '00 - {{IEEE}}. {{Global
  Telecommunications Conference}}. {{Conference Record}} ({{Cat}}.
  {{No}}.{{00CH37137}})}, vol.~2, Nov. 2000, pp. 1119--1123 vol.2.

\bibitem{tsunoda_supportingipleo_2004}
H.~Tsunoda, K.~Ohta, N.~Kato, and Y.~Nemoto, ``Supporting {{IP}}/{{LEO}}
  satellite networks by handover-independent {{IP}} mobility management,''
  \emph{IEEE Journal on Selected Areas in Communications}, vol.~22, no.~2, pp.
  300--307, Feb. 2004.

\bibitem{tsunoda_geographicalorbitalinformation_2006}
------, ``Geographical and {{Orbital Information Based Mobility Management}} to
  {{Overcome Last-Hop Ambiguity}} over {{IP}}/{{LEO Satellite Networks}},'' in
  \emph{2006 {{IEEE International Conference}} on {{Communications}}}, vol.~4,
  Jun. 2006, pp. 1849--1854.

\bibitem{roth_implementationgeographicalrouting_2021}
M.~M.~H. Roth, H.~Brandt, and H.~Bischl, ``Implementation of a geographical
  routing scheme for low {{Earth}} orbiting satellite constellations using
  intersatellite links,'' \emph{International Journal of Satellite
  Communications and Networking}, vol.~39, no.~1, pp. 92--107, 2021.

\bibitem{bertaux_softwaredefinednetworking_2015}
L.~Bertaux, S.~Medjiah, P.~Berthou, S.~Abdellatif, A.~Hakiri, P.~Gelard,
  F.~Planchou, and M.~Bruyere, ``Software defined networking and virtualization
  for broadband satellite networks,'' \emph{IEEE Communications Magazine},
  vol.~53, no.~3, pp. 54--60, Mar. 2015.

\bibitem{papa_dynamicsdncontroller_2018}
A.~Papa, T.~De~Cola, P.~Vizarreta, M.~He, C.~Mas~Machuca, and W.~Kellerer,
  ``Dynamic {{SDN Controller Placement}} in a {{LEO Constellation Satellite
  Network}},'' in \emph{2018 {{IEEE Global Communications Conference}}
  ({{GLOBECOM}})}, Dec. 2018, pp. 206--212.

\bibitem{papa_designevaluationreconfigurable_2020}
A.~Papa, T.~{de Cola}, P.~Vizarreta, M.~He, C.~{Mas-Machuca}, and W.~Kellerer,
  ``Design and {{Evaluation}} of {{Reconfigurable SDN LEO Constellations}},''
  \emph{IEEE Transactions on Network and Service Management}, vol.~17, no.~3,
  pp. 1432--1445, Sep. 2020.

\bibitem{walker_satelliteconstellations_1984}
J.~G. Walker, ``Satellite constellations,'' \emph{Journal of the British
  Interplanetary Society}, vol.~37, p. 559, 1984.

\bibitem{fossa_overviewiridiumlow_1998}
C.~Fossa, R.~Raines, G.~Gunsch, and M.~Temple, ``An overview of the {{IRIDIUM}}
  ({{R}}) low {{Earth}} orbit ({{LEO}}) satellite system,'' in
  \emph{Proceedings of the {{IEEE}} 1998 {{National Aerospace}} and
  {{Electronics Conference}}. {{NAECON}} 1998. {{Celebrating}} 50 {{Years}}
  ({{Cat}}. {{No}}.{{98CH36185}})}, Jul. 1998, pp. 152--159.

\bibitem{handley_delaynotoption_2018}
M.~Handley, ``Delay is {{Not}} an {{Option}}: {{Low Latency Routing}} in
  {{Space}},'' in \emph{Proceedings of the 17th {{ACM Workshop}} on {{Hot
  Topics}} in {{Networks}}}, ser. {{HotNets}} '18.\hskip 1em plus 0.5em minus
  0.4em\relax {New York, NY, USA}: {Association for Computing Machinery}, Nov.
  2018, pp. 85--91.

\bibitem{centerforinternationalearthscienceinformationnetwork-ciesin-columbiauniversity_griddedpopulationworld_2018}
{Center for International Earth Science Information Network - CIESIN - Columbia
  University}, ``Gridded {{Population}} of the {{World}}, {{Version}} 4
  ({{GPWv4}}): {{Population Density}}, {{Revision}} 11,'' {Palisades, New
  York}, 2018.

\bibitem{cartopy}
\BIBentryALTinterwordspacing
{Met Office}, \emph{Cartopy: a cartographic python library with a matplotlib
  interface}, Exeter, Devon, 2010 - 2015. [Online]. Available:
  \url{http://scitools.org.uk/cartopy}
\BIBentrySTDinterwordspacing

\bibitem{maihofer_surveygeocastrouting_2004}
C.~Maihofer, ``A survey of geocast routing protocols,'' \emph{IEEE
  Communications Surveys Tutorials}, vol.~6, no.~2, pp. 32--42, 2004.

\bibitem{alzaben_endtoendroutingsdn_2021}
N.~Alzaben and D.~W. Engels, ``End-to-{{End Routing}} in {{SDN Controllers
  Using Max-Flow Min-Cut Route Selection Algorithm}},'' in \emph{2021 23rd
  {{International Conference}} on {{Advanced Communication Technology}}
  ({{ICACT}})}, Feb. 2021, pp. 461--467.

\bibitem{eppstein_findingshortestpaths_1997}
D.~Eppstein, ``Finding the k {{Shortest Paths}},'' \emph{SIAM Journal on
  computing}, no. 28.2, p.~26, 1997.

\bibitem{little2008little}
J.~D. Little and S.~C. Graves, ``Little's law,'' in \emph{Building
  intuition}.\hskip 1em plus 0.5em minus 0.4em\relax Springer, 2008, pp.
  81--100.

\end{thebibliography}

\end{document}